# COMPUTATIONAL SIMULATION OF THERMAL PROCESSES IN NCS PRODUCTION PLANTS

Pablo Cortés, Omar López, Ramiro Medina, Gerardo Gordillo, Javier Guevara

**Abstract**

This paper develops a computational simulation of the fluid dynamics and heat transfer phenomena that occur inside combustion chamber, and flat and semispherical open evaporators used in NCS (non-centrifugal sugar) industry. The study estimates the effect of primary/secondary air ratio and excess air in the combustion chamber performance resulting in an optimal value of 1.3 for the equivalence ratio in the lower section of combustion chamber. The influence of both mechanisms of heat transfer in the evaporating and concentrating steps of the production process were studied (convection and radiation) as well as other parameters like geometric configuration, composition of heating gases and flux regimes. To provide a tool for designers, some correlations for Nusselt's number were developed. Several assumptions were made to reduce the computer's capacity required, like surface at constant temperature, properties of gases modelled by polynomials and negligible radiation at low temperatures. The mechanism of boiling in the liquid that is being heated was not studied. With this model, the authors found that a specific geometric configuration yielded the best results for evaporating and concentrating sugar in the flat-with-fins open evaporator.

INTRODUCTION

*1.1 Background*

Non centrifugal cane sugar (NCS) is the technical name given by FAO to the traditional product of concentrating cane juices by means of evaporation. This product is worldwide known with different names such as panela (LATAM), jaggery, piloncillo, raspadura, papelón, chancaca, among others. Colombia is the second world producer of panela with the largest per capita consumption [1]. The row material for NCS production is sugarcane, which has the second largest cultivated area in Colombia, making its waste material, sugarcane bagasse, an important source of biomass fuel. NCS is produced in a sugarcane mill, where the sugarcane is crushed to extract its juice. The remaining bagasse, known as green bagasse is burned in a furnace where the flue gases are used to provide heat for juice dehydration.

The production process of NCS starts with the extraction of sugarcane juices using a milling machine. Juice represents the 60% of sugarcane total weight [2]. The extracted juice is then clarified increasing its temperture to 90°C and adding clarifyng substances at 65°C, in order to remove impurities. This step is performed using flat open evaporators.

The next step in NCS production is the evaporation process, during which the juice is concentrated from 17 to 92 °Brix using semi-spherical open evaporators. The final product is then molded to its desired shape and cooled down by natural convection. A general overview of the production process is given by [3], and a schematic map of the hole process is shown in Illustration 1.

Six key areas of research and development have been identified for the NCS industry [4] in Colombia:

1. Contamination and environmental impact
2. Deforestation
3. Energy efficiency and recovery
4. Juice losses during milling process
5. Low quality final product
6. Empirical and local development of production techniques

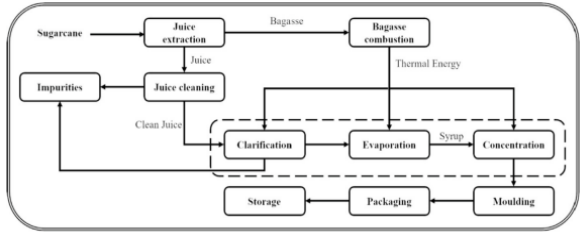

*Illustration 1. NCS production process [3]*

Few research has been carried on NCS furnaces, and on the heat transfer processes and equipment used in NCS production. The aim of this article is to present an extensive study on the thermal processes of a conventional NCS production plant, specifically on combustion and heat transfer processes.

*1.2 Organization*

This paper is organized in 5 sections. The second section is dedicated to study the combustion process on a NCS production plant furnace. Section 3 and 4 are dedicated to study the convection and radiation process that occur between the combustion gases and underside of the open evaporators. Finally, section 5 presents the main results and conclusions.

*1.3 General design of combustion chamber and open evaporators in NCS production plants*

Illustration 2 shows the three main thermal processes that occur in a typical NCS production plant. First, green bagasse is burnt in the combustion chamber in order to generate the necessary thermal energy for the hole process. Then, the combustion gases are conducted through a duct where they transfer their energy content to the cane juices by means of flat and semi-spherical open evaporators (2,4,5 and 6).

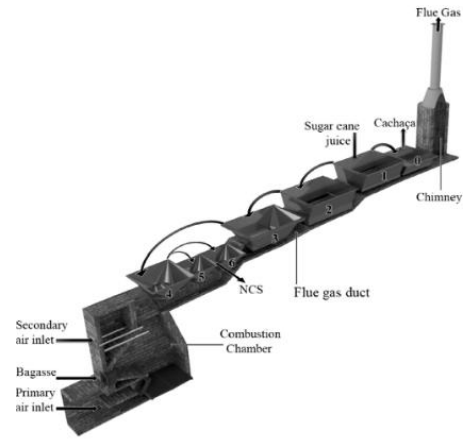

*Illustration 2.Thermal processes in NCS production: (0-1) clarification, (2-4) evaporation, (5-6) concentration. Taken from [3]*

The latest and most efficient design for green bagasse is known as WARD-CIMPA (Illustration 3), which incorporates independent inlets for primary air, secondary air and bagasse, together with a drying sub-chamber that allows burning high moisture bagasse.

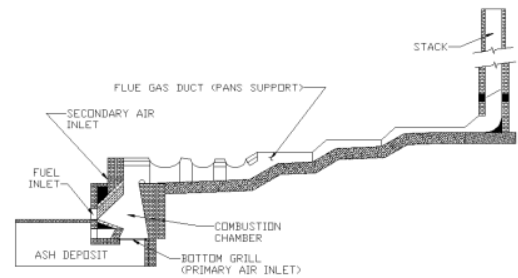

*Illustration 3. Furnace with WARD-CIMPA combustion chamber*

Up to date, there have been experiments and mathematical models of sugar cane furnaces [5], [6] that characterize the global combustion process with no details about distribution of chemical reactions, species and temperatures in the combustion chamber; also there have been CFD simulations with a domain covering only flue gas duct and stack [7], and CFD models of combustion chambers with similar geometry but burning different kind of fuels [8] and not considering the drying process of the fuel on the drying sub-chamber, nor the independent entries of primary and secondary air [9]. As a consequence, a CFD model of the combustion chamber in NCS furnaces is an important contribution to improve furnace design and operation procedures in order to maximize furnace efficiency and minimize environmental emissions.

On the other hand, the thermal design of flat and semi-spherical open evaporators has been carried out using Nusselt's correlations for heat transfer process in pipes and around spheres respectively [10]. Illustration 4 represents the most common designs for flat and semi-spherical open evaporators.

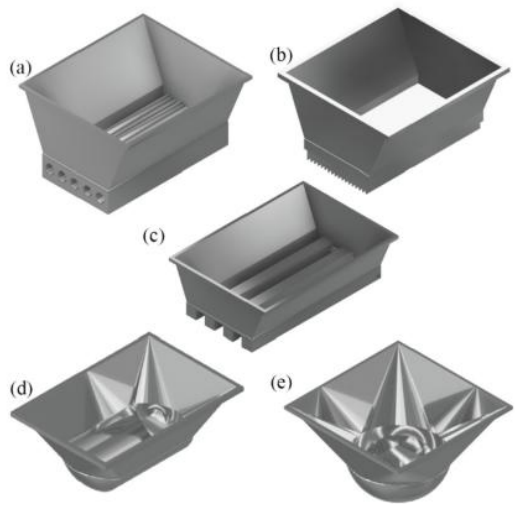

*Illustration 4. Common designs of flat and semi-spherical open evaporators used in NCS industry. a) pyro-tubular, b) flat with fins, c) grooved, d) semi-cylindrical, e) semi-spherical. Image taken from [3]*

The study of heat transfer processes on both semy-spherical and flat open-evaporators is an important contribution for designers as long as there is not an exact correlation for Nusselt's number in the considered geometries.

COMBUSTION CHAMBER

## 2.1 *Proposed model*

The model consists of a combustion chamber from a furnace where 487 kg/h of green sugar cane with a moisture content of 50% are burned and the hot flue gases are used to dehydrate sugar cane juice and produce 160 kg/h of brown sugar cane. The combustion chamber, a WARD-CIMPA type, consists of two connected sections. The lower section, which receives primary air from below and green sugar cane from a side door, acts as an up-draft (counter-flow) fixed bed gasifier. The upper section receives hot gases from the gasification process, which are further combusted with the addition of secondary air. In the lower section, primary air enters at atmospheric pressure and room temperature by a grill located in the bottom over which is resting the bed. Green bagasse is fed at room temperature by lots through a side door at regular time intervals. The bed surface is an open zone where the produced gases converge and then pass to upper section of chamber. The char content of the sugar cane is partially combusted in the lower section of the chamber. This combustion provides heat for other processes taking place in the bed, such as heating, drying, pyrolysis, and gasification of the remaining char.

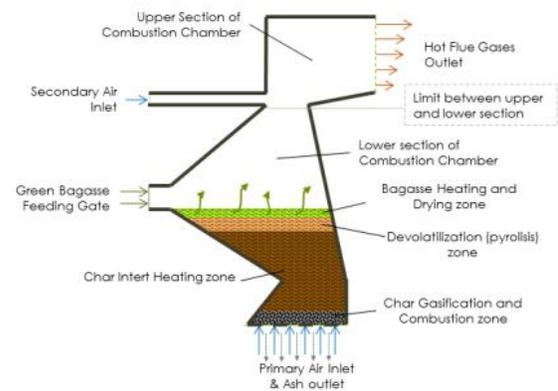

*Illustration 5. Modeled Ward-CIMPA combustion chamber.*

According to [11], char gasification and combustion zones overlap, and both take place in a narrow layer in the bed bottom, while the heating and drying of bagasse is a rapid process in a narrow layer on top of bed. The pyrolysis zone will be also located next to the top of the bed and the remaining bed section will comprise a char inert heating zone (Figure 2). The mesh size was selected by a convergence analysis using cold air to solve flow distribution in the chamber until velocity magnitude at flue gas outlet didn't change with a higher cell number A block structured mesh was finally selected with 186,461 quadrilateral cells. The model domain boundaries were defined as combustion chamber walls, inlets of primary air, secondary air and fuel, and the beginning of semi-spherical open-evaporator 4 in Illustration 2. Gas mix properties such as mass diffusion, viscosity, and thermal conductivity are derived from [12], 1990 and were introduced to the model by an user defined function:

$$D_{mix} = 1.78 \times 10^{-5} (T_g/300)^{1.75} \ [m^2/s] \quad (1)$$
$$\mu_{mix} = 1.98 \times 10^{-5} (T_g/300)^{2/3} \ [N \cdot s/m^2] \quad (2)$$
$$\lambda_{mix} = 4.8 \times 10^{-4} T_g^{0.717} \ [W/m \cdot K] \quad (3)$$

## 2.2 Simplifications adopted.

In order to reduce the computational complexity of the proposed model, the following simplifications were adopted:

1. Bi-Dimensional model.
2. Stationary state.
3. Adiabatic and no slip walls.
4. Pressure loss through bed is estimated considering bed as a porous media with constant porosity.
5. Bed material is modelled as spheres with constant diameter.
6. Non reacting ash.
7. Laminar flow through the bed.

Assumption 4 is widely used in gasification process models [11], and assumption 5 is recommended by [13] when no further information is available. Combustion chamber model was implemented in commercial software ANSYS FLUENT 14 as a species transport problem with finite rate chemistry. Solid particles were represented using DPM (Discrete Phase Model). Mass and energy exchange with gas phase are calculated by DPM model, while momentum interaction is calculated as momentum source terms using the porous media approach.

## 2.3 Characterization of green bagasse

Green Sugarcane Bagasse characterization was taken from [14] (see Table 1). The empirical formula of green sugarcane bagasse was obtained using ultimate analysis together with atomic weights of each element, then chemical equation was balanced to determine the stoichiometric air required.

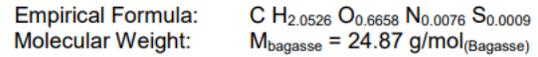

Empirical Formula: $CH_{2.0526}O_{0.6658}N_{0.0076}S_{0.0009}$
Molecular Weight: $M_{bagasse} = 24.87 \ g/mol_{(Bagasse)}$

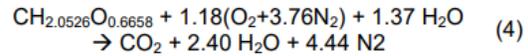

$$CH_{2.0526}O_{0.6658} + 1.18(O_2 + 3.76N_2) + 1.37 \ H_2O \rightarrow CO_2 + 2.40 \ H_2O + 4.44 \ N_2 \quad (4)$$

| | | As Received | Dry Basis | DAF |
|---|---|---|---|---|
| Moisture Content | MC | 50.00% | 0 | 0 |
| Volatile Matter | VM | 41.25% | 82.50% | 83.67% |
| Fixed Carbon | FC | 8.05% | 16.10% | 16.33% |
| Ash | Hash | 0.70% | 1.40% | 0 |
| | H2O | 50.00% | 0 | 0 |
| | Ash | 0.70% | 1.40% | 0 |
| | %C | 23.81% | 47.62% | 48.30% |
| | %H | 4.10% | 8.20% | 8.32% |
| | %O | 21.12% | 42.24% | 42.84% |
| | %N | 0.21% | 0.42% | 0.43% |
| | %S | 0.05% | 0.11% | 0.11% |
| HHV (kJ/kg) | | 8700 | | 17975 |

Table 1. Characterization of sugar cane bagasse

## 2.4 Composition of volatile matter

The estimation of the composition of volatile matter released during pyrolysis, was done with software CEA (Chemical Equilibrium and Species). This package determines equilibrium composition and molar fraction of more than 150 possible species for a given fuel composition, temperature, and pressure. According to the thermo-gravimetric analysis (TGA) of sugar cane bagasse developed by

[14], the pyrolysis takes place between 500 and 700K, so the latter temperature and atmospheric pressure were used in the model to determine volatile composition (Table 2).

| Species | Molar Fraction |
|---|---|
| $CH_4$ | 0.1496 |
| CO | 0.0035 |
| $CO_2$ | 0.0830 |
| $H_2$ | 0.1014 |
| $H_2O$ | 0.2574 |
| C | 0.4051 |

Table 2. Volatile composition at 700K

### 2.5 Porous media definition

Sugarcane bagasse size distribution expressed as diameter of equivalent spheres [15] was used to estimate an average equivalent diameter for bagasse particles (Table 3).

| Size range ($\emptyset_{p,eq}$) | Mass fraction (Y) |
|---|---|
| < 10μm | 0.65 |
| 10 – 50 μm | 2.03 |
| 50 – 100 μm | 2.64 |
| 100 – 500 μm | 25.81 |
| 500 – 1000 μm | 32.34 |
| 1000 – 5000 μm | 34.78 |
| 5000 – 10000 μm | 1.65 |
| > 10000 μm | 0.1 |

Table 3. Sugarcane bagasse size distribution

According to [16], when determining an average particle diameter from a size distribution chart, volume-area average ($\emptyset_{p,eq,v-a}$) gives better results when representing momentum interaction between solid and gas phases, while area-volume average ($\emptyset_{p,eq,a-v}$) gives better results when representing mass transfer between phases such as reaction rates.

$$\emptyset_{p,eq,v-a} = \frac{\sum_{i=1}^{n} \emptyset_{p,i}^{2} Y_i}{\sum_{i=1}^{n} \emptyset_{p,i}^{3} Y_i} = 4.01 mm \quad (5)$$

$$\emptyset_{p,eq,a-v} = \frac{1}{\sum_{i=1}^{n} \frac{Y_i}{\emptyset_{p,i}}} = 0.267 mm \quad (6)$$

The volume-area average diameter is then used to calculate viscous resistance ($1/\alpha$) and inertial resistance ($C_2$) from the correlations derived [17].

$$\alpha = \frac{\emptyset_p^2}{150} \frac{\varepsilon^3}{(1-\varepsilon)^2} \quad , \quad C_2 = \frac{3.5}{\emptyset_p} \frac{(1-\varepsilon)}{\varepsilon^3} \quad (7),(8)$$

The area-volume average diameter is used for particle injections and the user defined function (UDF) implemented for heterogeneous reaction rate calculation. Apparent density of bagasse fed is 127 kg/m3 and porosity is 0.6 [18], so particle density at bed surface is 317.5 kg/m3 and will decrease while approaching the bottom of the bed.

### 2.6 Evolution of the bagasse particles

A bagasse particle evolves from green bagasse to ash while traveling through the combustion chamber; particle's evolution stages are shown in Illustration 6

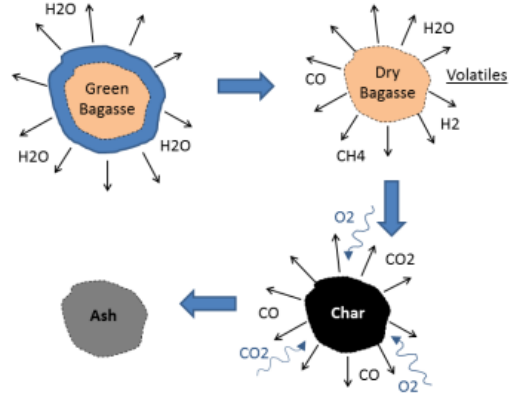

Illustration 6. Evolution of bagasse particles

### 2.6.1 Initial inert heating

Takes place when green sugarcane bagasse is fed to combustion chamber at ambient temperature and then heated to moisture evaporation. During this process there is no mass exchange, and energy exchange is given by:

$$m_p Cp_p \frac{dT_p}{dt} = hA_p(T_g - T_p) + \varepsilon_p A_p \sigma (T_R^4 - T_p^4) \quad (9)$$

Inert heating/cooling processes also take place from the end of drying to the start of pyrolysis, from the end of pyrolysis to start of char combustion, and from the end of combustion until ash particle leaves the model domain.

### 2.6.2 Drying

Bagasse moisture content is modeled as a volume fraction of liquid water added to the particle surface. The drying process consists of the evaporation of bagasse moisture content when particle temperature reach moisture evaporating temperature, and bagasse moisture boiling when particle temperature reach moisture boiling temperature and continues while particle mass is greater than initial mass of dry particle. Mass transferred between phases is equal to the weight of water evaporated/boiled by the heat received by the particle; energy transferred is equal to evaporating latent heat of such amount of water during evaporation and boiling, and the sensible heat due to temperature change during evaporation. Mass and energy exchange during moisture evaporation:

$$-\frac{dm_p}{dt} = k_m A_p \rho_g \ln\left(1 + \frac{Y_{i,s} - Y_{i,g}}{1 - Y_{i,s}}\right) \quad (10)$$

$$m_p Cp_p \frac{dT_p}{dt} = hA_p(T_g - T_p) - \frac{dm_p}{dt} h_{fg} + \varepsilon_p A_p \sigma(T_R^4 - T_p^4) \quad (11)$$

Mass and energy exchange during moisture boiling:

$$-\frac{dm_p}{dt} h_{fg} = hA_p(T_g - T_p) + \varepsilon_p A_p \sigma(T_R^4 - T_p^4) \quad (12)$$

FLUENT doesn't consider condensation of vapor when its partial pressure reaches water saturation pressure which take place when water vapor rises through the bed and find cold particles of green bagasse on its way. Because of this, it was introduced an UDF that define a new DPM law which re-condenses excess water vapor on particle surface when relative humidity exceeds 100%, and transfers condensation latent heat and mass from vapor to particle.

### 2.6.3 Devolatilization (Pyrolysis)

During this process, dry bagasse particle is decomposed into a mix of gases and char (carbon and ash) by the action of temperature. The gases mix composition was estimated using CEA. This process takes place when particle temperature reaches pyrolysis start temperature and particle mass is greater than its initial mass of char. Pyrolysis start temperature was estimated around 500K in Thermo-gravimetric Analysis performed by [14], where it was also determined that leading mechanism for devolatilization kinetics of sugarcane bagasse, is a first order reaction:

$$-\frac{dm_p}{dt} = k_j(m_p - (\%_m char)m_{p,initial}) \quad (13)$$

$$k_j = Ae^{\left(-\frac{E}{RT}\right)} \quad (14)$$

Heat and mass transferred between phases during pyrolysis is determined according to:

$$m_p Cp_p \frac{dT_p}{dt} = hA_p(T_g - T_p) - \frac{dm_p}{dt} \Delta H_{pyrolysis} + \varepsilon_p A_p \sigma(T_R^4 - T_p^4) \quad (15)$$

Where $\Delta H_{pyrolysis}$ is the latent heat of devolatilized species (pyrolysis enthalpy of reaction), according to [11] this enthalpy is near zero and can be neglected. Pyrolysis is then modeled as a single global reaction:

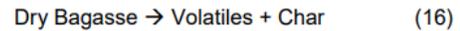
Dry Bagasse → Volatiles + Char  (16)

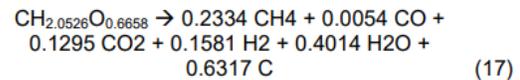
$CH_{2.0526}O_{0.6658}$ → 0.2334 $CH_4$ + 0.0054 CO + 0.1295 $CO_2$ + 0.1581 $H_2$ + 0.4014 $H_2O$ + 0.6317 C  (17)

### 2.6.4 Particle combustion

Particle combustion process starts when particle releases all its moisture and volatile content and continues while particle mass is greater than their initial content of ash. There are many models available to describe particle surface reaction kinetics, in this case it was selected an exposed unreacted core model in which reaction rate considers gas film diffusion resistance together with chemical reaction kinetics, with no consideration of resistance through ash layer. This model was selected because of the low ash content of bagasse [11].

$$\frac{dm_p}{dt} = \frac{M_{char}}{M_i} \frac{6(1-\varepsilon)Y_{i,g}\rho_g \forall_g}{\varphi \emptyset_p} \left(\frac{1}{\frac{1}{k_j} + \frac{1}{k_m}}\right) \quad (18)$$

Chemical reaction rate ($k_j$) is described by an Arrhenius equation:

$$k_j = A_j T^{\beta_j} e^{\left(\frac{-E_j}{RT}\right)} \quad (19)$$

Mass and energy interaction between phases is defined according to:

$$m_p C p_p \frac{dT_p}{dt} = hA_p(T_g - T_p) - f_h \frac{dm_p}{dt} H_{comb} + \varepsilon_p A_p \sigma (T_R^4 - T_p^4) \quad (20)$$

where $f_h$ is the fraction of energy from chemical reaction that is absorbed by particle.

2.7 *Movement of particles in the bed*

There is no previous information available regarding the bed zones where the interaction processes between gas and solid phases occur. As a result, instead of decomposing the bagasse feeding into individual sources of moisture, volatiles, carbon, and ash, with each one injected in a different zone of the bed, as done by [19], we assumed a uniform flow of bagasse from the bed surface to the bottom grill, with each particle following a predetermined path and evolving based on the gas composition and temperatures encountered along the way. Given the simplifications used in the model regarding constant porosity and particle diameter, the volumetric flow of bagasse will remain constant, and the vertical velocity of particles at a given location will be a function of the cross-sectional area of the furnace. Once the vertical velocity is calculated, the horizontal velocity is determined to keep the particle on its pre-defined path (see Illustration 7).

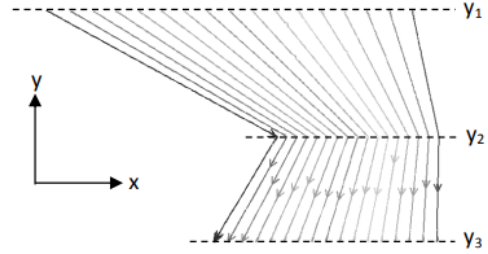

*Illustration 7. Predefined paths of particles in the bed*

2.8 *Carbon oxidizing reaction*

Carbon oxidation is a heterogeneous reaction in which carbon oxidizes to CO and $CO_2$ according to

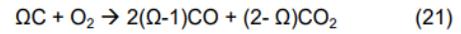

$$\Omega C + O_2 \rightarrow 2(\Omega-1)CO + (2-\Omega)CO_2 \quad (21)$$

It was selected the correlation from [20] for $CO/CO_2$ ratio and temperature, given its capacity to predict $CO/CO_2$ ratios at high temperatures [21]

$$\Omega = 2\left[\frac{1 + 4.3e^{(-3390/T_p)}}{2 + 4.3e^{(-3390/T_p)}}\right] \quad (22)$$

The total disappearance rate of C on particle surface (r1) is calculated using a UDF (User Defined Function in FLUENT) where is also calculated the parameter $\Omega$ according to particle surface temperature. Carbon oxidizing reaction was included in the model as two independent reactions: oxidation to $CO_2$ and oxidation to CO. The software calculates heat and mass transfer between phases for each reaction according to calculated reaction rate (r1) and parameter $\Omega$.

2.9 *Chemical reactions*

Chemical reaction included in the model, together with their kinetics constants are described in Table 4. Reactions ID 1 to ID 3 are heterogeneous while reactions ID 4 to ID 9 are homogeneous.

| ID | Nombre | Equation | ΔH$_R$ | A | E [J/kmol] | Reaction Rate $r_i$ |
|---|---|---|---|---|---|---|
|  | combustion-1 | $\Omega C + O_2 \rightarrow 2(\Omega-1)CO + (2-\Omega)CO_2$ |  | 1.715 T$_p$ | 7.483x10$^7$ | $r_1 = \frac{M_C}{M_{O2}} \frac{6(1-\varepsilon)Y_{O2}\rho_g \forall_g}{\varphi \emptyset_p} \left( \frac{1}{\frac{1}{k_1}+\frac{1}{k_m}} \right)$ |
| 1 | combustion-1a | $C + O_2 \rightarrow CO_2$ | -32750 [kJ/kg$_C$] |  |  | $r_1[2(\Omega-1)/\Omega]$ |
| 2 | combustion-1b | $C + ½ O_2 \rightarrow CO$ | -9250 [kJ/kg$_C$] |  |  | $r_1[(2-\Omega)/\Omega]$ |
| 3 | gasification | $C + CO_2 \rightarrow 2CO$ | 14250 [kJ/kg$_C$] | 3.42 T$_p$ | 1.297x10$^8$ | $\frac{M_C}{M_{CO2}} \frac{6(1-\varepsilon)Y_{CO2}\rho_g \forall_g}{\varphi \emptyset_p} \left( \frac{1}{\frac{1}{k_3}+\frac{1}{k_m}} \right)$ |
| 4 | pyrolysis | Volatiles → aCH4 + bCO + cCO2 + dH2 + eH2O | 0 | 5x10$^{22}$ | 2.67x10$^8$ | $k_4(m_p - (\%_m char)m_{p,0})$ |
| 5 | combustion-2 | $CO + ½O_2 \rightarrow CO_2$ | -10107 [kJ/kg$_{CO}$] | 2.239 x10$^{12}$ | 1.7 x10$^8$ | $\varepsilon k_5 C_{CO} C_{O2}^{0.25} C_{H2O}^{0.5}$ |
| 6 | dissociation | $CO_2 \rightarrow CO + ½O_2$ | 6410 [kJ/kg$_{CO2}$] | 5x10$^8$ | 1.7x10$^8$ | $\varepsilon k_6 C_{CO2}$ |
| 7 | combustion-3 | $CH_4 + 1.5O_2 \rightarrow CO + 2H_2O$ | -32375 [kJ/kg$_{CH4}$] | 5.012 x10$^{11}$ | 2x10$^8$ | $\varepsilon k_7 C_{CH4}^{0.7} C_{O2}^{0.8}$ |
| 8 | combustion-4 | $H_2 + ½O_2 \rightarrow H_2O$ | -120500 [kJ/kg$_H$] | 9.87x10$^8$ | 3.1x10$^7$ | $\varepsilon k_8 C_{H2} C_{O2}$ |
| 9 | water-gas-shift | $CO + H_2O \leftrightarrow CO_2 + H_2$ | -1470 [kJ/kg$_{CO}$] | 2780 | 1.258x10$^7$ | $\varepsilon k_9 \left( C_{CO} C_{H2O} - \frac{C_{CO} C_{H2O}}{k_r} \right)$ |

*Table 4. Chemical reactions considered in the model.*

## 2.10 Radiation model

Radiation was included in the simulation using P-1 model which is the simplest case of the P-N model. The radiation energy source in conservation equation is calculated as the divergence of the radiation flux.

$$S_R = -\nabla \cdot q_r = -4\pi \left( an^2 \frac{\sigma T^4}{\pi} + E_p \right) - (a + a_p)G \quad (23)$$

Where Ep is the equivalent emission of the particles, ap is the equivalent absorption coefficient, n is the refractive index of the medium and G is the incident radiation. P-1 radiation model is only valid for porosity values near 1 when is used together with discrete phase model, otherwise it will consider all particles in the bed are receiving radiation even if the particle is located behind another particle, in order to avoid this as bed porosity is 0.6, it was included an UDF that only enables radiation model for the particles located in the top of the bed, while the heat transfer due to radiation of particles located below the top is included in the heat transfer by conduction, using an effective thermal conductivity ($\lambda_s^*$)

$$\lambda_s^* = \varepsilon \lambda_{rg} + \varepsilon \frac{\lambda_s}{\left[ \frac{\lambda_s}{d_p \lambda_{rs}} + 1.43(1-1.2\varepsilon) \right]} \quad (24)$$

$$\lambda_{rg} = 4 \sigma \, 0.05 \, T_g^3 \quad , \quad \lambda_{rs} = 4 \sigma \, 0.85 \, T_s^3 \quad (25),(26)$$

$$\lambda_s = 1.3 \times 10^{-3} + 5 \times 10^{-5} T + 6.3 \times 10^{-7} T^2 \quad (27)$$

## 2.11 Results

The model was solved for eight (8) cases composed of four PA flows (four ER values at bed) and two EA values. Cases considered are shown in Table 6. Mass weighted averages of temperature and species mass percentage at inlet of pan No. 1 are shown in Table 5. For every case analyzed, the model performed similarly with a few differences that will be pointed next. Char combustion and gasification take place in a narrow layer at the bottom of the bed, as predicted by [11]. The primary air entering from the bottom grill is preheated and then supply oxygen for char combustion where carbon is primarily oxidized to CO2 with just a few portion oxidized directly to CO; CO2 formed is immediately converted to CO by gasification and dissociation reactions (ID 2 and 6 respectively), then the resulting CO, N2 and O2, continue rising through an inert thick layer until bed top where devolatilization and drying take place in another narrow layer.

| Case No. | 1 | 2 | 3 | 4 | 5 | 6 | 7 | 8 |
|---|---|---|---|---|---|---|---|---|
| Primary air flow [kg/s] | 0.3544 | 0.3544 | 0.2953 | 0.2953 | 0.2531 | 0.2531 | 0.2215 | 0.2215 |
| Percentage of total air flow [%] | 80.0% | 66.7% | 66.7% | 55.6% | 57.1% | 47.6% | 50.0% | 41.7% |
| Equivalence relation at lower section of chamber | 1.25 | 1.25 | 1.5 | 1.5 | 1.75 | 1.75 | 2.00 | 2.00 |
| Secondary air flow [kg/s] | 0.0886 | 0.1772 | 0.1477 | 0.2363 | 0.1899 | 0.2785 | 0.2215 | 0.3101 |
| Percentage of total air flow [%] | 20.0% | 33.3% | 33.3% | 44.4% | 42.9% | 52.4% | 50.0% | 58.3% |
| Total air flow [kg/s] | 0.4430 | 0.5316 | 0.4430 | 0.5316 | 0.4430 | 0.5316 | 0.4430 | 0.5316 |
| Excess Air (EA) [%] | 0% | 20% | 0% | 20% | 0% | 20% | 0% | 20% |
| Primary air / Secondary air Ratio (PA/SA) | 4.00 | 2.00 | 2.00 | 1.25 | 1.33 | 0.91 | 1.00 | 0.71 |
| Equivalence relation (ER) for char combustion | 0.66 | 0.66 | 0.79 | 0.79 | 0.92 | 0.92 | 1.06 | 1.06 |

*Table 6. Cases considered.*

| Caso | 1 | 2 | 3 | 4 | 5 | 6 | 7 | 8 |
|---|---|---|---|---|---|---|---|---|
| AP/AS | 4.00 | 2.00 | 2.00 | 1.25 | 1.33 | 0.91 | 1.00 | 0.71 |
| T (K) | 1150 | 1121 | 1119 | 1071 | 1063 | 1035 | 1016 | 966 |
| %$_{mass}$ $CH_4$ | 0.15% | 0.14% | 0.49% | 0.37% | 0.87% | 0.85% | 0.96% | 0.59% |
| %$_{mass}$ $O_2$ | 1.04% | 3.10% | 2.19% | 5.09% | 3.76% | 5.40% | 5.29% | 7.79% |
| %$_{mass}$ $CO_2$ | 18.81% | 18.97% | 19.67% | 16.22% | 16.92% | 17.95% | 17.52% | 14.69% |
| %$_{mass}$ CO | 1.52% | 1.35% | 1.72% | 1.34% | 1.81% | 1.70% | 1.92% | 1.45% |
| %$_{mass}$ $H_2O$ | 19.52% | 18.23% | 18.55% | 18.86% | 17.63% | 16.04% | 16.16% | 16.90% |
| %$_{mass}$ $H_2$ | 0.02% | 0.02% | 0.04% | 0.03% | 0.08% | 0.06% | 0.12% | 0.08% |
| %$_{mass}$ $N_2$ | 58.94% | 58.19% | 57.34% | 58.08% | 58.92% | 58.00% | 58.04% | 58.51% |

*Table 5. Temperature and species at inlet of semi-spherical open evaporator No1*

Combustion of volatiles only takes place at the bed top, as predicted by [22]. The geometry of the combustion chamber causes hot gases and remaining oxygen after char combustion and gasification to be distributed mainly by the bed's side opposite to the feeding gate, while the gate side receives little flow for heating and volatiles combustion. As a result, the feeding gate side reaches lower temperatures and higher concentrations of CH4, H2, and H2O than the opposite side. The temperature distribution in the lower section of the chamber can be divided into two sections: in the bed and above the bed. The cases modeled consider a primary air flow higher than required for the oxidation of char. Therefore, the bed temperature will decrease as the air flow increases since there is a fixed amount of char to release heat during oxidation. All additional air must be heated and, therefore, will decrease the bed temperature. On the other hand, the cases modeled consider reducing atmospheres in the lower section of the combustion chamber (ER>1). Thus, the air that did not react with char will oxidize volatiles above the bed, and any additional air will allow more volatiles to be oxidized. It was observed that the heat released during volatiles combustion compensates for the lower temperatures achieved in the bed when increasing the primary air flow. Therefore, as the primary air increases, gas temperatures leaving the lower section of the chamber also increase, given that the equivalence ratio in the lower section of the chamber remains above 1. Part of the heat from volatiles combustion is transferred to particles in the top of the bed, making these particles hotter than particles in the char combustion/gasification zone. Temperature profiles for a given particle path vary with EA values. As an example, Illustration 8 shows temperature profiles for particle path 540 of 684.

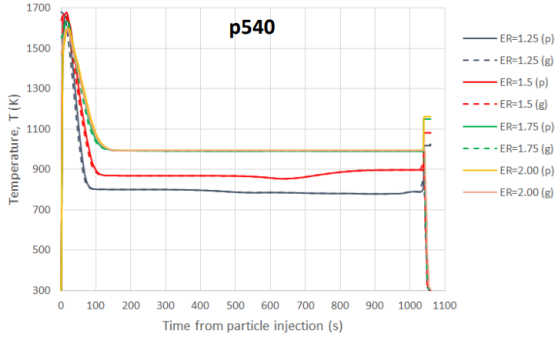

*Illustration 8. Bed temperature profiles for a given particle path.*

Illustration 9 shows the relationship between mass weighted average of gas temperature at inlet of pan No. 1 and PA/SA ratio, for 0% and 20% EA. For 0% EA, maximum gas temperature is achieved for a PA/SA ratio of 3.25 equivalent to a 1.31 ER at lower section of chamber, while for 20% EA maximum gas temperature is achieved for a PA/SA ratio of 1.88 equivalent to a 1.28 ER at lower section of camera.

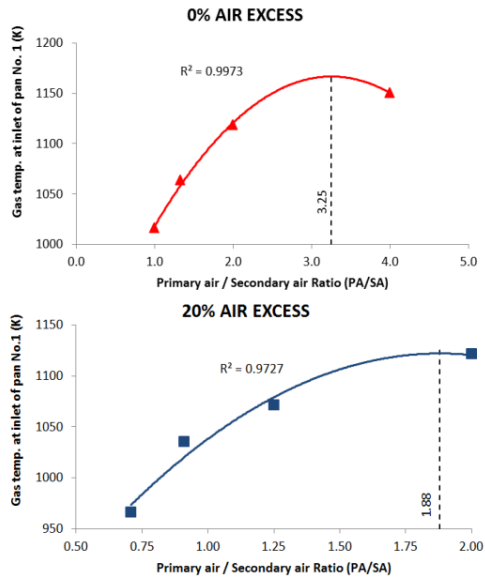

*Illustration 9. Effect of PA/Sa ratio on flue gas temperature*

Illustration 10 shows the relationship between mass weighted average of species mass percentage at inlet of pan No. 1 and PA/SA ratio, for 0% and 20% EA. As PA/SA ratio approaches its optimum value (maximum flue gas temperature) for each EA, combustible species content (CO, $CH_4$, $H_2$) decreases, and resulting species ($CO_2$, $H_2O$) increases. When increasing EA, $O_2$ increases while combustible species slightly decreases.

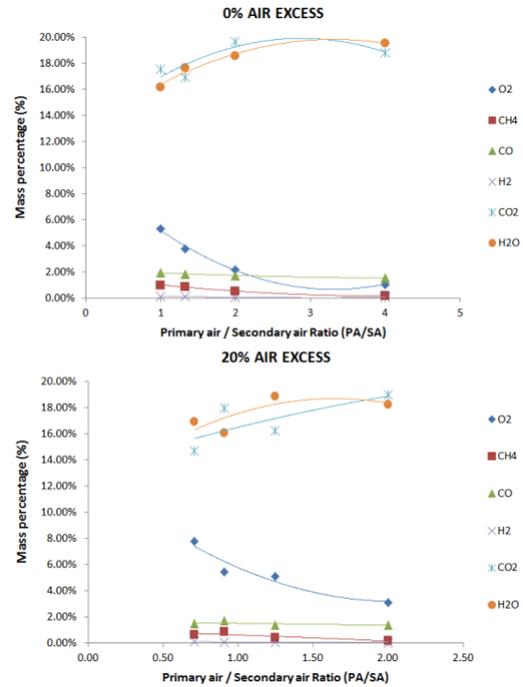

*Illustration 10 Effect of PA/SA ratio on flue gas composition*

Distribution of parameters along cross-section of duct at the inlet of pan No. 1 is not uniform; there is a clearly visible gas path with higher velocities, temperatures, and combustible gas concentrations; the zone above this path is rich in oxygen with lower temperatures; the zone below this path is a recirculation zone with low temperatures and velocities near zero, in the opposite direction of the main flow (Illustration 11).

In the lower section of chamber, peak temperature is located above bed top in the side opposed to feeding gate where the flow of hot gases and remaining oxygen from char combustion and gasification is higher. Peak temperature in upper section of chamber is located next to SA inlet where injected oxygen completes combustion of volatiles. Illustration 12 shows behavior of peak temperatures as ER in the lower section of chamber and global EA varies.

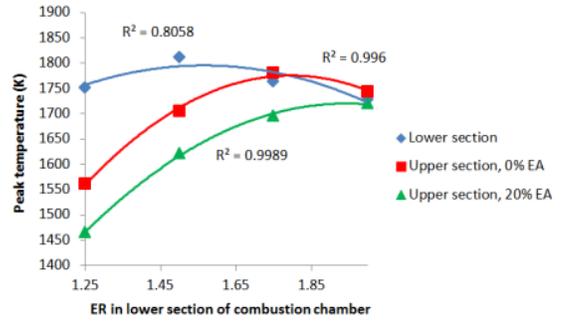

*Illustration 12 Peak temperatures in combustion chamber*

It can be seen that when increasing global EA, gas volume to be heated increases and thus flue gas temperature is reduced, but higher oxygen content and secondary air velocity improve mixing of air with unreacted volatiles which reduces combusting species at flue gas outlet, and improves combustion efficiency; this can be verified when comparing flue gas temperature to adiabatic equilibrium temperature estimated by CEA as seen in Illustration 12, where the flue gas temperature for cases with 20% EA are closer to their adiabatic equilibrium temperature than cases with 0% EA.

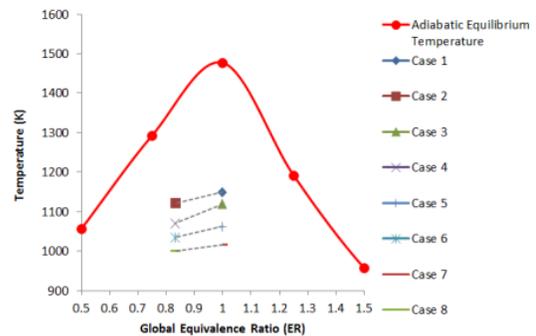

*Illustration 13. Comparison between flue gases temperatures and estimated adiabatic equilibrium temperature.*

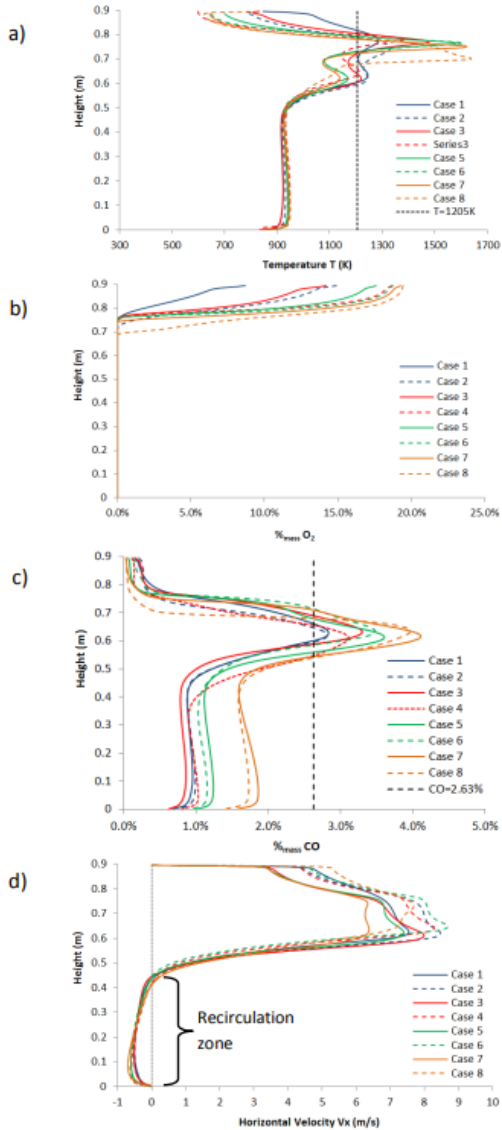

*Illustration 11 Cross section distribution at inlet of pan No 1. A) Temperature, B). %$_{mass}$ $O_2$, C). horizontal velocity, D). %$_{mass}$ CO*

When bed ER increases, peak temp. at upper section tends to be higher than peak temp. at lower section, but none of the simulated cases exceeds the 1800 K threshold, when $NO_x$ formation rate becomes important.

### 2.11.1 *Comparison of results to real data*

Real data from existing furnaces are scarce, the only available data is the design data from of an existing furnace, performed in a zero-dimensional model and claimed to be representative of actual furnace performance.

According to this information, the flue gases temperature entering to pan No. 1 is 932°C (1205 K), and mass percentage of CO at flue gas outlet is 2.63%, which validates temperatures range estimated by model, while CO mass content tends to be under predicted by the model with values 27 to 50% lower when compared to mass weighted average, but consistent with peak values estimated when compared to cross section distribution. See Illustration 11 a and c.

SEMI-SPHERICAL OPEN EVAPORATORS

3.1 *Proposed model*

*3.1.1 Geometry*

The problem was developed in 3 dimensions due to the variable and complex geometry of the cross-section of the duct as the combustion gases move forward. The relevant dimensions for generating the semi-spherical evaporator geometry (illustration 14) are as follows.

- Duct width: 0.7 meters
- Evaporator diameter: 1.3 meters
- Evaporator height: 0.5 meters
- Duct's height after the last semi-spherical evaporator: 0.6 meters

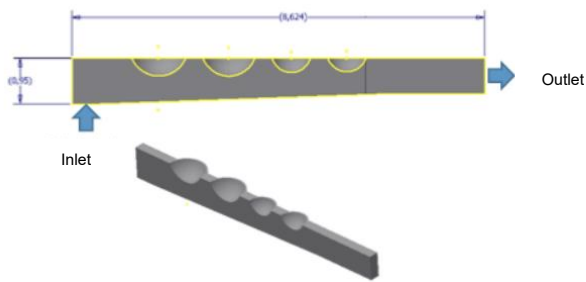

*illustration 14. Duct's geometry at the section of semi-spherical evaporators*

*3.1.2 Discretization of the computational domain*

The mesh used to discretize the computational domain was hybrid, composed of tetrahedral and prismatic elements. 10 layers of prisms were generated on the walls with a growth rate of 1.1, followed by tetrahedral elements. Densities were generated over the region that includes each evaporator to capture both the flow of gases and heat transfer with high accuracy. A mesh convergence analysis was performed as follows: Three meshes were generated: a fine one (13 million elements), a medium-sized one (10 million elements), and a coarse one (8 million elements), where the convergence of the solution for dry air at 950°C will be verified, corresponding to the temperature of the combustion gases' inlet in the region of the semi-spherical evaporators.

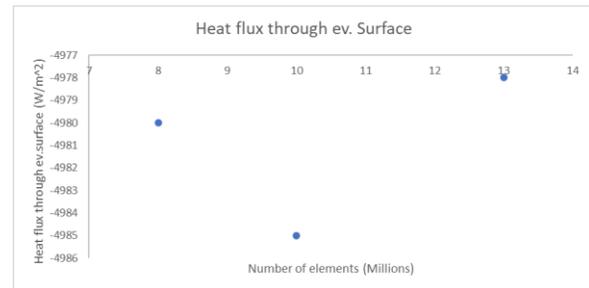

*Illustration 15. Convergence analysis for semi-spherical open evaporator meshes.*

Illustration 15 shows that even for the coarse mesh, the computational problem has converged in terms of the elements number.

*3.1.3 Solver configuration*

Navier-Stokes equations of fluid motion are solved in the discretized domain using commercial software Fluent v.15. The solver configuration was set up as follows:

- Steady state
- Incompressible flow motion
- Transport equations: Mass, momentum, energy, and species
- Newtonian fluid
- Solution scheme: SIMPLE
- Turbulence model: Standard k-ε
- Convergence criteria: absolute

The boundary conditions for momentum and mass equations applied to the computational domain are shown in illustration 16. Boundary conditions for energy equation are shown in illustration 17. Temperature at evaporators' surface was set to 100°C since this is the boiling point of NCS.

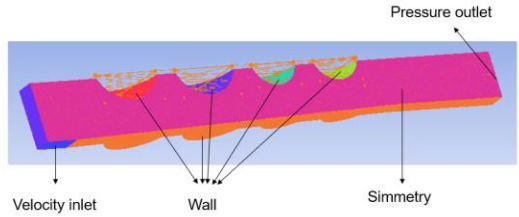

illustration 16. Boundary conditions for semi-spherical evaporators' computational domain.

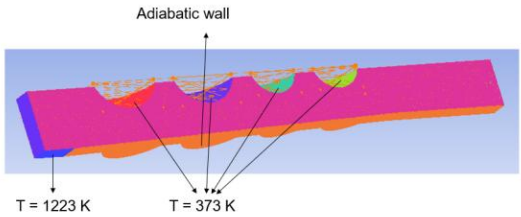

illustration 17. Boundary conditions for semi-spherical evaporators' computational domain.

First, the convection model was implemented without radiation to analyze the correlation of the Nusselt number with Reynolds and Prandtl numbers, which corresponds to a purely convective analysis. After this, the radiative model was implemented to analyze the contribution of radiation to the first semi-spherical evaporator inlet, where the highest temperatures of the combustion gases are present.

### 3.1.4 Nusselt correlation estimation

The simulations were carried out according to a $3^2$ factorial design protocol: 3 levels and 2 factors (Reynolds' and Prandtl's numbers). Therefore, a total of 9 simulations were performed; based on these results, a model to relate Nusselt's with Reynold's and Prandtl's number is proposed with the following general form:

$$Nu = C * Re^m * Pr^m \qquad (28)$$

Where

- Nu: Nusselt's number
- Re: Reynolds' number
- Pr: Prandtl's number
- C, m, n: Unknown constants

To find Nusselt's number, it is necessary to know the convective heat transfer coefficient (h) in order to obtain a correlation for each semi-spherical evaporator. The equation that correlates the convective coefficient with heat flow is presented below:

$$h = \frac{q_w}{T_w - T_f} \qquad (29)$$

where

- h: Convective heat transfer coefficient
- $T_w$: evaporator's wall temperature
- $T_f$: Fluid's reference temperature
- $q_w$: Heat flux through evaporator's wall

Levels of both Reynolds and Prandtl's numbers used in the $3^2$ experiment are shown in Table 7. The three levels of Prandtl's number corresponds to air, gases mixture (as predicted by the combustion chamber simulation) and 20% air excess (as predicted by the combustion chamber simulation).

| Factor | Levels | | |
|---|---|---|---|
| Re | 7000 | 17000 | 27000 |
| Pr | 0.726 (air) | 0.755 (mixture) | 0.751 (air excess) |

Table 7. Levels for the two factors considered in the $3^2$ experiment.

### 3.1.5 Radiation model

Additionally to the analysis of convective heat transfer, an analysis of heat transfer due to radiation solely was performed. The Discrete Ordinates (DO) model included in Ansys Fluent was used, as it corresponds to the most robust method for radiation analysis. The simulation was performed for Re=17000 and Pr of the gas mixture.

### 3.2 Results

The ANOVA analysis of the $3^2$ experiment showed that Prandtl's number is not significant for the levels considered in this study. In Illustration 18, the response surface of the experiment is show.

higher at this point, especially in the area where the gases impact.

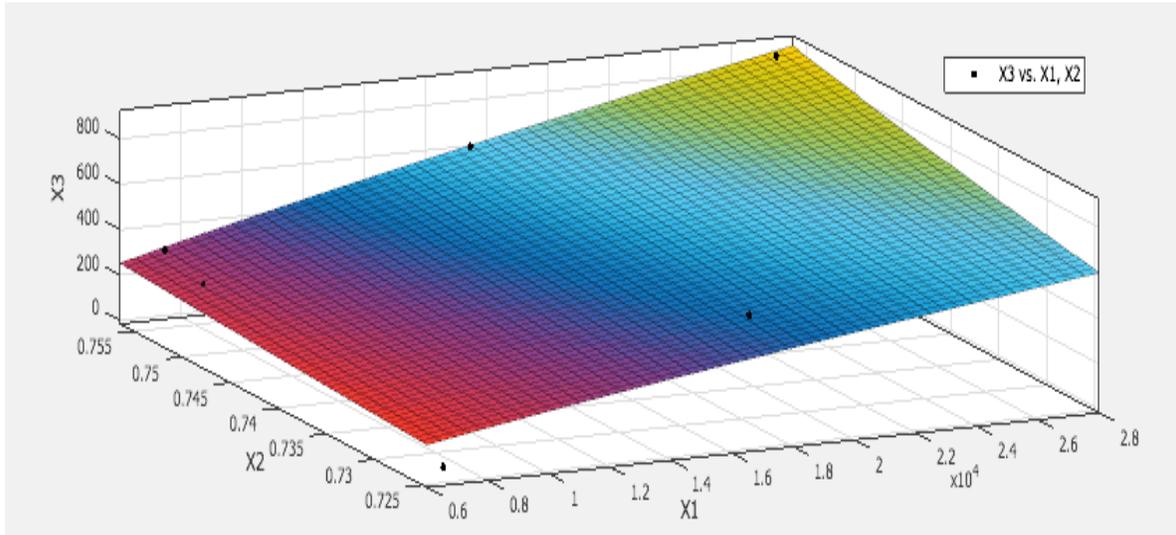

*Illustration 18. Surface response for the $3^2$ experiment considered for the convective study. X1=Reynolds, X2=Prandtl, X3=Nusselt*

Therefore, the following relations were developed for Nusselt's number, only as a function of Reynold's number:

- Air:
  $Nu = 0.4189 * Re^{0.771}$   (30)
- Mixture (as predicted by combustion chamber simulation):
  $Nu = 0.6391 * Re^{0.743}$   (31)
- Air excess 20% (as predicted by combustion chamber simulation):
  $Nu = 0.5149 * Re^{0.743}$   (32)

Besides, as Prandtl's number does not have a significant effect on the results for the levels considered in this study, in the following it will be shown only the results for Pr=0.726, that is Prandtl's number of air. In Illustration 19, it is shown the shape and the position of the planes used to plot contours of velocity and temperature.

As it is shown in Illustration 20, the temperature difference is very high in the vicinity of the symmetry plane, since in this region the gases impact the surface perpendicularly. For this reason, heat flux is

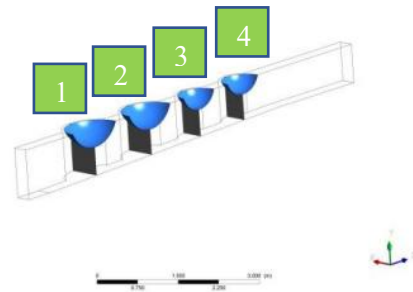

*Illustration 19. Position of planes used to plot velocity and temperature contours. Numbers correspond to open semi-spherical evaporators.*

Regarding the upper left part of the image, it is evident that there is a slight temperature increase at this point because of trapping zones caused by the geometry of the furnace at the boundaries.

Regarding the successive evaporators (Illustration 20, Illustration 21, Illustration 22, Illustration 23), it is observed that the temperature difference over the symmetry line decreases considerably, because of the heat transmission that takes place in the previous evaporators from the gases to sugar cane juices. It is also evident from the temperature contours of evaporator 3 evaporator 4 that the temperature remains at its highest magnitude around the center of the gas duct. This is because the area/volume ratio at the top of the duct is high, so the gases lose heat more rapidly at this point towards the sugar cane juice.

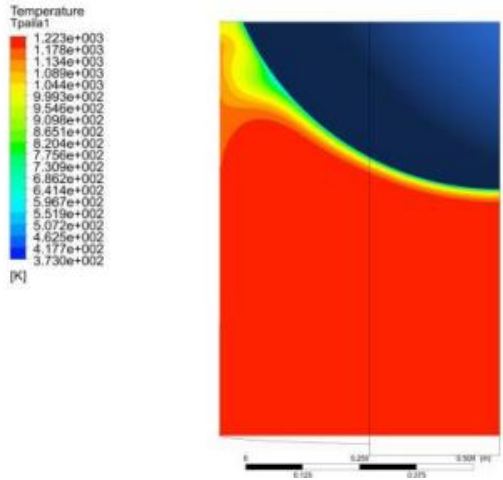

*Illustration 20. Temperature contour in the zone below open-evaporator number 1*

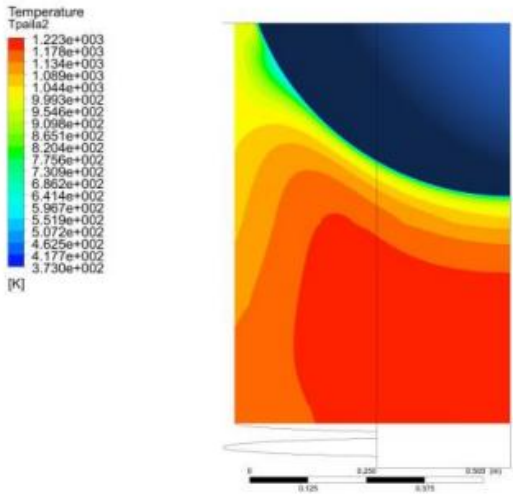

*Illustration 21. Temperature contour in the zone below open semi-spherical evaporator number 2.*

The flow lines shown in Illustration 24 justify what was explained above regarding the high temperature differences observed in the zone below the semi-spherical evaporator 1. It is evident that the flow lines impact the lower area of the first pan, causing a large amount of heat transfer and high temperature in that area.

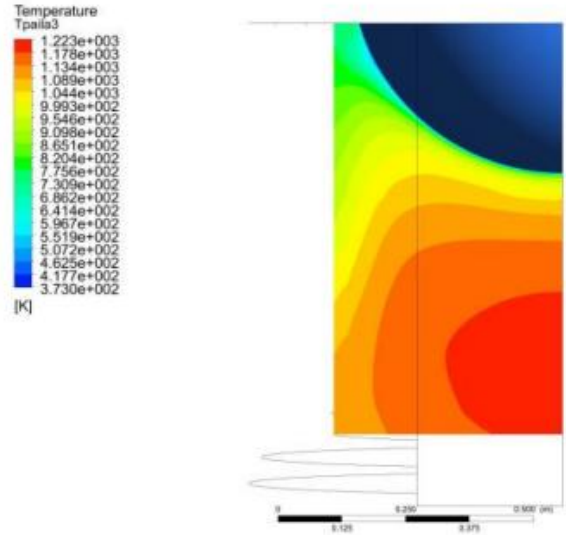

*Illustration 22. Temperature contour in the zone below open-evaporator number 1*

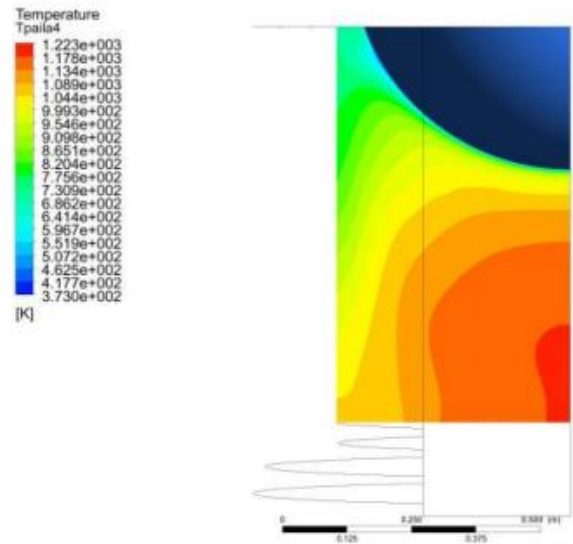

*Illustration 23. Temperature contour in the zone below open-evaporator number 1*

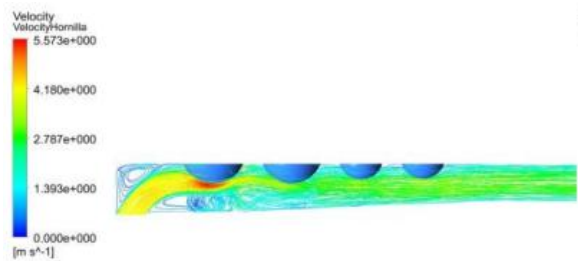

*Illustration 24. Velocity contour plotted on symmetry plane.*

Thanks to the curved shape of duct's boundaries, recirculation zones and abrupt changes in the trajectory of the particles are generated when entering those zones. It is shown in Illustration 25.

symmetry plane to the limit line of the duct. Therefore, a low magnitude of velocity is observed as the flow lines follow the trajectory adopted from semi-spherical evaporator number 1.

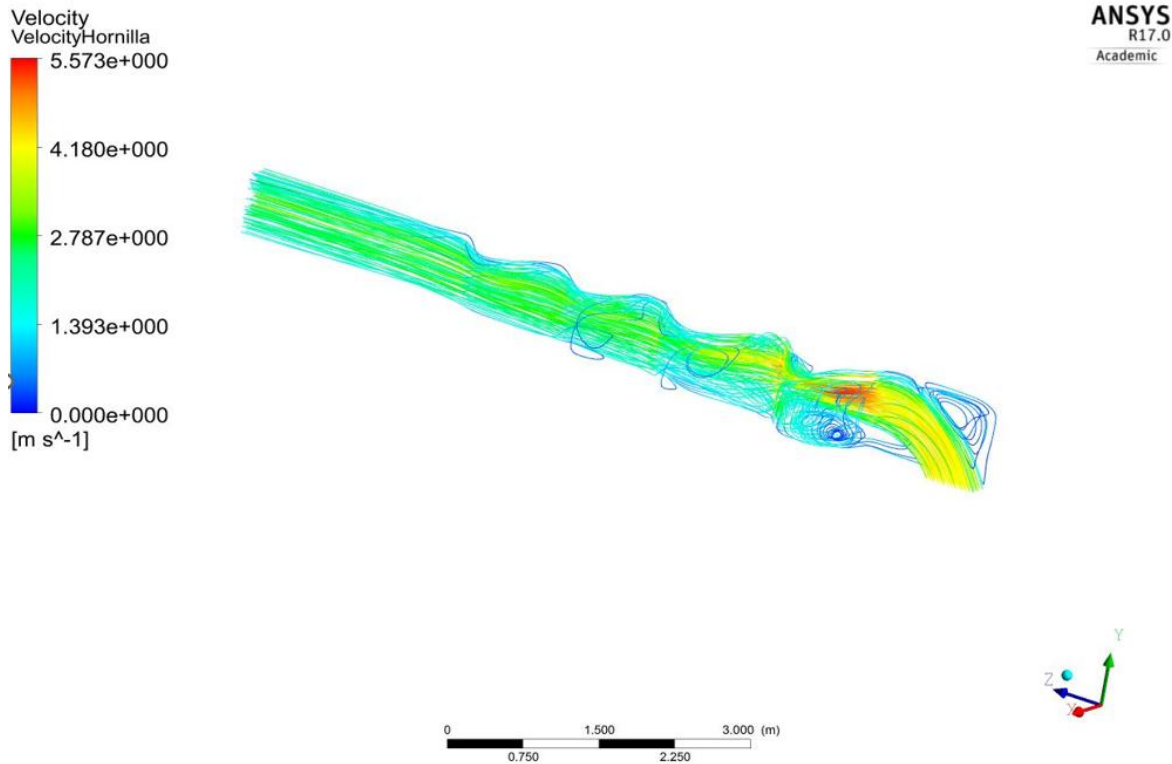

*Illustration 25. Flow lines plotted on the duct's wall.*

Considering the phenomenon studied with the temperature contours previously shown, the hypotheses can be corroborated through an analysis of velocity contours over the same regions, where very low or zero velocities occur. This suggest that there are significant pressure losses in this zone due to the intricate geometry that the flow lines must pass through. As shown in Illustration 14, recirculation occurs as the air impacts the semi-spherical evaporator and, due to the magnitude of its velocity and the geometry of duct's walls, its trajectory is changed.

It is evident from the velocity contours of the evaporators number 2, number 3 and number 4, that a considerable drop in velocity is produced near the duct's walls. This is because the main trajectory of gases goes from the

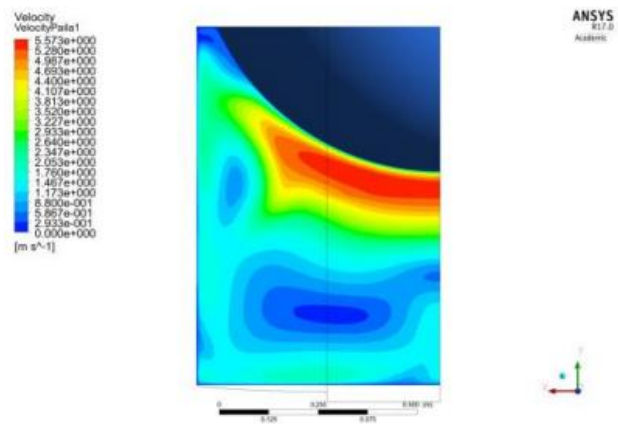

*Illustration 26. Velocity contours in the zone below the open semi-spherical evaporator number 1*

The results for the radiation model are shown in Table 8. They correspond to a simulation made for Re=17000 and Pr for gas mixture (as predicted by combustion chamber simulation).

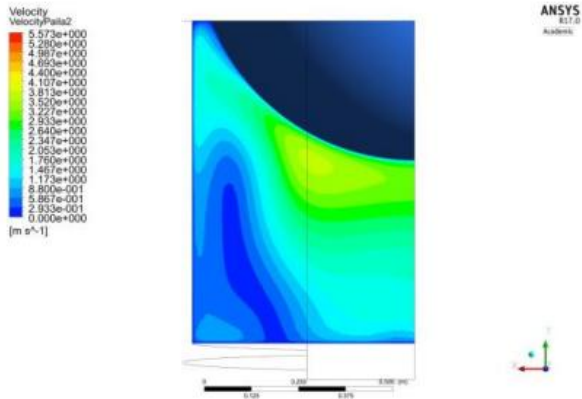

*Illustration 27. Velocity contours in the zone below the open semi-spherical evaporator number 2*

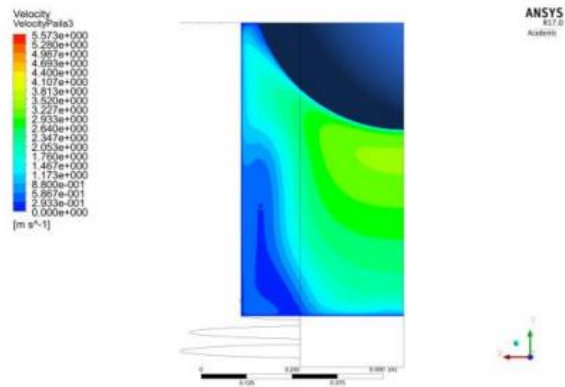

*Illustration 28. Velocity contours in the zone below the open semi-spherical evaporator number 3*

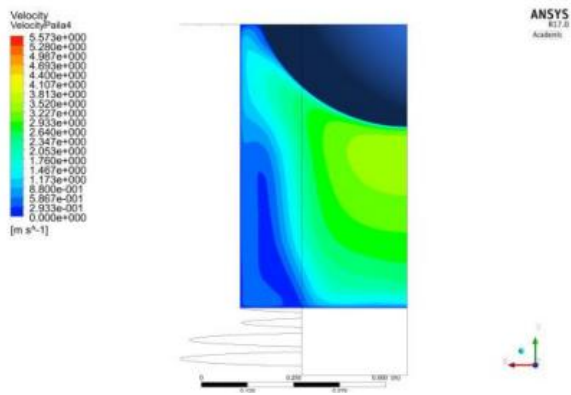

*Illustration 29. Velocity contours in the zone below the open semi-spherical evaporator number 4*

|  | Heat transfer due to radiation (W/m2) | Total heat transfer (W/m2) | Radiation contribution % |
|---|---|---|---|
| Open evap. 1 | 24271 | 39379 | 61,63 |
| Open evap. 2 | 13303 | 24191 | 54,99 |
| Open evap. 3 | 10021 | 18415 | 54,42 |
| Open evap. 4 | 10230 | 17983 | 56,89 |

*Table 8. Simulation results for radiation heat transfer*

### 3.3 *Comparison with experimental results*

[23] performed an experimental analysis to determine heat transfer coefficient in both semi-spherical and flat with fins open evaporators. The experimental results showed a participation of radiation with respect to the total heat transfer for open evaporator number 1 of 64%, compared with the value of 61.63% gotten from the simulations. Similarly, the experiment result for the total heat transfer reported was 30500 W/m$^2$, meanwhile the simulation result was 39379 W/m$^2$.

### 4. FLAT WITH FINS OPEN EVAPORATOR

### 4.1.1 Geometry

In Illustration 30 is shown a schematic view of the duct in the zone where flat with fins open evaporators are installed. The model considered in this study has the following geometrical parameters:

- Fins' length: 10 cm
- Fins' width: ¼ in
- Duct's width: 1.1 m
- Duct's height: 0.42
- Flow conditions: Re=21000, T_gas=973K, T_wall=373K

These parameters are typical of this kind of facilities and are reported in [24], where a design of a NCS production plant is carried out. [24] also choose 5 cm as the gap among fins. With the aim of studying the behavior of heat transfer as a function of this gap, three different values were considered: 5cm, 2.5 cm and 1 cm. The geometry was generated using a commercial CAD software. Shows a view of the computational domain set up for this study.

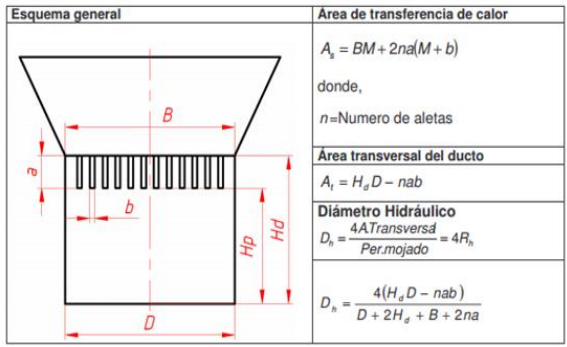

*Illustration 30. Schematic view of the duct in the zone below flat with fins evaporator. Taken from [24]*

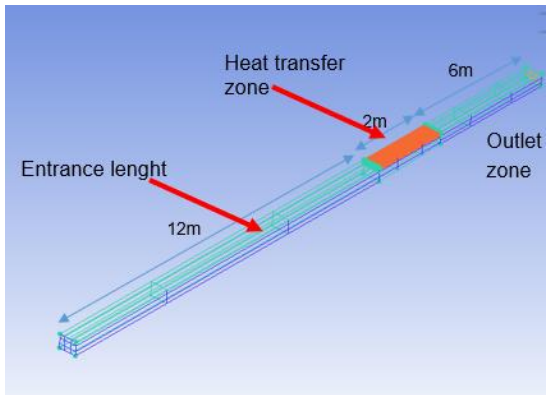

*illustration 31. Geometry to solve the flat with fins open evaporator problem.*

### 4.1.2 Discretization of computational domain

Due to the parallelepiped characteristic of the geometry, a set of structured meshes were created to discretize the computational domain. The set of meshes had the following number of elements

- Coarse mesh: 1.5 million elements
- Gross mesh: 3 million elements
- Medium mesh: 4.5 million elements
- Fine mesh: 6 million elements
- Extra fine mesh: 7.5 million elements

The height of the first element was set up to 0.1 mm, and the growing rate for the next elements was 1.1.

A convergence analysis was performed on the heat flux. In, the results of this analysis are displayed in illustration 32. With 6 million elements, the heat flux has converged, and in consequence, the fine mesh, with 6 million elements was chosen.

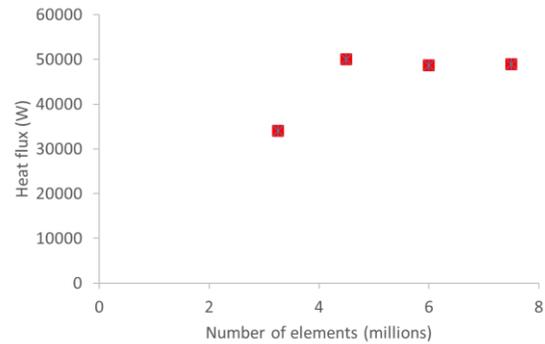

*illustration 32. Convergence analysis for flat with fins open evaporator.*

### 4.1.3 Solver configuration

As in the semi-spherical open evaporator, Navier-Stokes' equations of flow motion are the key to simulate the thermal behavior foe this kind of devices. Using the fine mesh generated before, commercial software FLUENT was used to solve Navier-Stokes' equations numerically. The configuration of FLUENT's solver was:

- Steady state
- Solution scheme: SIMPLE
- Solver: Pressure based
- Convergence criteria: Absolute
- Transport equations: Mass, momentum, energy
- Turbulence model: Standard k-ε

Boundary conditions used for momentum and energy equations are displayed in illustration 33. The conditions at the inlet zone were taken from [24]. In this study, only convection was considered, since, due to the duct's geometry in the zone below flat with fins open evaporators, radiation heat transfer can be easily calculated by designers using the well-known correlations for view factors.

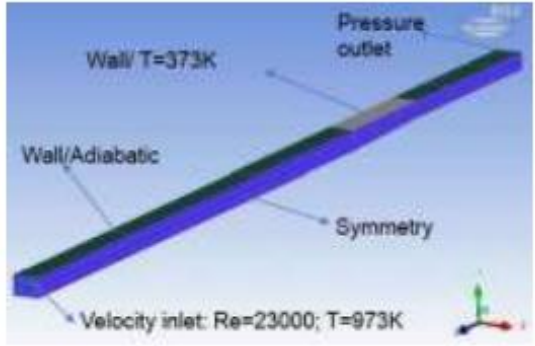

*illustration 33. Boundary conditions for both momentum and energy equations.*

To determine the heat transfer coefficient due to convection, equation (29) is considered. However, a more precise way to estimate the convective heat transfer coefficient can be implemented by varying the surface's temperature of the evaporator. Then

$$q_{w1} = h(T_{w1} - T_f) \quad (33)$$

And

$$q_{w2} = h(T_{w2} - T_f) \quad (34)$$

If difference among $T_{w1}$ y $T_{w2}$ is not so large, the convective heat transfer coefficient will be approximately constant, and

$$h = \frac{dq}{d(\Delta T)} \quad (35)$$

Reynold's number was varied over the interval [12000,30000]. To select the number of cases to simulate, the methodology descripted by [25] was followed.

### 4.2 Results

Illustration 34 shows the results of heat flux for four different temperatures, Re=21000 and spacing between fins of 5 cm (7th case of Table 9). The slope of the line formed by the four simulated pints in Illustration 34 corresponds to the convective heat transfer coefficient; for this case this value was found to be 5.2 W/m²K. The linear behavior of the response is similar for all cases of Table 9, and are omitted here for brevity.

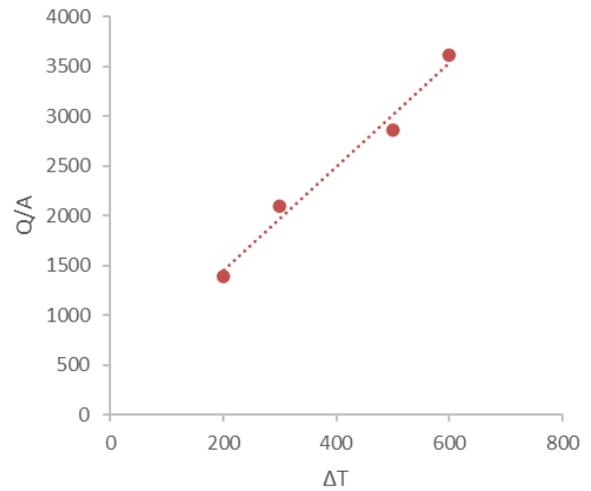

*Illustration 34. Simulation results for Re=21000, spacing between fins of 5cm and air. 7th case of Table 9*

Besides, as it was stated before, it is of particular interest to find out the optimal spacing between fins. Illustration 35 shows the influence of the spacing between fins on the convective heat transfer coefficient, for spacings of 5, 2.5 and 1 cm, at Re of 21000. It is clear the behavior of convective heat transfer coefficient as a function of the spacing between fins is not linear, and that it has a maximum somewhere near 3.5 cm of spacing.

| Reynold's number | Fluid | | | | | |
|---|---|---|---|---|---|---|
| | Air | | | Mixture (as predicted by combustion chamber simulation) | | |
| | Spacing between fins | | | Spacing between fins | | |
| | 5 cm | 2.5 cm | 1 cm | 5 cm | 2.5 cm | 1 cm |
| 12000 | 1 | 2 | 3 | 4 | 5 | 6 |
| 21000 | 7 | 8 | 9 | 10 | 11 | 12 |
| 30000 | 13 | 14 | 15 | 16 | 17 | 18 |

*Table 9. Simulation cases for flat with fins open evaporator.*

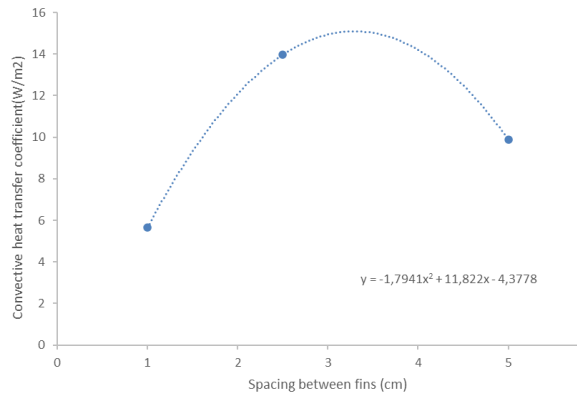

*Illustration 35. Effect of spacing between fins on convective heat transfer coefficient (cases 7th, 8th, and 9th of Table 9)*

As it was found for semi-spherical open evaporators, the influence of Prandtl's number on the convective heat transfer coefficient in the interval chosen is not significant. Consequently, a relation depending only on Reynold's number was developed to estimate Nusselt's number for spacing between fins of 5 cm.

$$Nu = 0{,}0282 * Re^{0{,}83} \qquad (36)$$

A similar relation was also developed for a spacing between fins of 2.5 cm:

$$Nu = 0{,}044 * Re^{0{,}987} \qquad (37)$$

4.3 *Comparison with experimental results*

As stated before, [23] performed an experimental analysis for both semi spherical and flat with fins evaporators. For flat with fins open evaporator reported a heat flux due to convection of 7 2651W/m². For the same condition, the simulation result gotten with the model previously described was 3615 W/m², with a difference of 36%. This difference could be due to the simplifications and assumptions made to run the simulations, and to the experimental procedure that averages the temperature in the zone below the evaporators.

5. *Conclusions*

Three of the main thermal processes of an NCS production plant were simulated. WARD-CIMPA combustion chamber geometry considers a larger bed surface for receiving heat by radiation during drying process, but also promotes preferential flows through bed which causes uneven temperature distribution. According to this, it is suggested that lower section of combustion chamber becomes symmetric with bottom grill optimized to ensure gas is uniformly distributed at bed top. Mixing rate between volatiles coming from lower section of chamber and SA injected should be improved as there are combustion reactions even beyond evaporator No. 1. According to this, it is suggested to evaluate new configurations for SA injection such as prolonged ducts directly to zones with higher concentrations of combustible species or additional injection ports in the opposite location of existing ones. Another solution for this mixing rate issue is to extend flue gas duct and move away pans, thus allowing more time to complete volatiles combustion before transferring heat to pans.

Regarding the design of the gas duct and the shape of the semi spherical open evaporators, it was found that they are susceptible to be improved in the following ways: firstly, the interface between the semi spherical open evaporators and the duct wall has abrupt changes in geometry, which generates a change in the trajectory of the gases. In this sense, it is necessary to round the edges of the interface, as well as to increase the area/volume ratio on the top of the evaporator. Simultaneously, recirculation occurs in the gas entry region, which must be alleviated by modifying the geometry of the walls so that it follows the trajectory of the gases. Lastly, as can be seen in the relationship found for the convective phenomenon, the Nusselt number turns out to be highly dependent on the Reynold's number, and Prandtl factor is not significant for the interval studied.

Regarding the design of the flat with fins open evaporator, it was found that the optimal spacing between fins is 3.5cm. With this spacing, the evaporator's effectiveness is significantly increased, which would eventually allow for a reduction in its length. It was also possible to obtain correlations for the Nusselt number as a function of the Reynolds number, for both types of evaporators which could be used in the design phase of a NCS plant.